\documentclass[aps,amsmath,amssymb,pre,superscriptaddress,twocolumn]{revtex4}
\usepackage{graphicx}
\usepackage{epsfig}
\usepackage{amsmath}
\usepackage{amssymb}

\begin{document}

\title{In-plane force fields and elastic properties of graphene }

\author{G.~Kalosakas}
\affiliation{Materials Science Department, University of Patras,
Rio GR-26504, Greece}
\affiliation{ICE-HT/FORTH, PO Box 1414, GR-26504 Rio, Greece}

\author{N.~N.~Lathiotakis}
\affiliation{Theoretical and Physical Chemistry Institute, NHRF, Vass. Constantinou 48,
Athens GR-11635, Greece}

\author{C.~Galiotis}
\affiliation{Materials Science Department, University of Patras,
Rio GR-26504, Greece}
\affiliation{ICE-HT/FORTH, PO Box 1414, GR-26504 Rio, Greece}

\author{K.~Papagelis}
\affiliation{Materials Science Department, University of Patras,
Rio GR-26504, Greece}
\affiliation{ICE-HT/FORTH, PO Box 1414, GR-26504 Rio, Greece}


\begin{abstract}

Bond stretching and angle bending force fields, appropriate to describe in-plane motion
of graphene sheets, are derived using first principles' methods.
The obtained force fields are fitted by analytical anharmonic energy potential functions, providing efficient
means of calculations in molecular mechanics simulations.
Numerical results regarding the mechanical behavior of graphene monolayers under various loads,
like uniaxial tension, hydrostatic tension, and shear stress, are presented, using both
molecular dynamics simulations and first principles' methods.
Stress-strain curves and elastic constants, such as, Young modulus, Poisson ratio,
bulk modulus, and shear modulus, are calculated.
Our results are compared with corresponding theoretical calculations as well as
with available experimental estimates.
Finally, the effect of the anharmonicity of the extracted potentials on the mechanical properties
of graphene are discussed.

\end{abstract}

\maketitle

\section{Introduction  \label{intro} }

Graphene consists of a two-dimensional ($2D$) sheet of covalently bonded carbon and forms
the basis of both $1D$ carbon nanotubes, $3D$ graphite but also of important commercial products,
such as, polycrystalline carbon (graphite) fibres. As a defect-free material, graphene is predicted to have
an intrinsic tensile strength higher than any other known material \cite{Zhao} and tensile stiffness similar
to values measured for graphite. Graphene over the years and even prior to its isolation has been an ideal
material to model as far as its mechanical properties are concerned. In particular, the elastic moduli of
single layer graphene and its elastic response, have been a subject of intensive 
theoretical research in recent years and quite different approaches have been employed
\cite{liu, Cadelano, fasolino,zhaoNL, Huang, Ni, Pour, Tsai, Kudin, Odegard, Meo}.
For example, several groups have performed first principles' calculations \cite{liu}, other
used empirical potentials for atomistic simulations \cite{Huang,fasolino,zhaoNL},
and also tight-binding methods have been employed \cite{Cadelano, zhaoNL}.
As it is evident from the literature survey there is a large discrepancy of values regarding
the stiffness of monolayer graphene and values ranging from 0.5 $TPa$
to 4 $TPa$ have been proposed depending on the methodology pursued in each case.

From the experimental point of view,  recent experiments have indeed confirmed the extreme stiffness
of graphene of 1 $TPa$ and provided an indication of the breaking strength of graphene of 42 $N/m$
(or 130 $GPa$ considering the thickness of graphene as 0.335 $nm$) \cite{lee}. These experiments
involved the simple bending of a tiny flake by an indenter on an AFM set-up and the force-displacement
response was approximated by considering graphene as a clamped circular membrane made by an isotropic
material. An alternative way to assess how effective graphene is in the uptake of applied stress or strain
(uniaxial or biaxial) is to probe the vibrational characteristics of certain interatomic bonds upon loading.
In particular, Raman spectroscopy has been proved very successful in monitoring the mechanical response
of certain Raman active modes upon the application of external mechanical stress/strain. In axial tension,
a linear relationships between Raman frequency and strain was established regardless of the geometry
of the monolayer graphene up to strains of about 1.5 \% \cite{Tsouk, OF1, OF2,zabel}.

In this work, we present appropriate empirical force fields, derived from first principles'
calculations,
to describe bond stretching and angle bending interactions in graphene.
Analytical nonlinear potentials are provided for these interactions, in order to efficiently
implemented  in atomistic molecular dynamics or Monte Carlo simulations.
A big advantage of atomistic models compared to the more accurate first principles' methods 
is their computational efficiency which makes them more practical for molecular dynamics
simulations especially for finite temperatures. 
Moreover, when analytical potentials are available, the effect of various types of interactions
can be investigated by switching off, or modifying, corresponding potential energy terms,
being able to improve in this way our understanding of the studied problem~\cite{KNF}.
The obtained empirical potentials are then used to calculate, through molecular
dynamics simulations, the mechanical response of graphene under various loads.
To test these results, density functional theory calculations of graphene's elastic behavior have been also performed.

The derived bond stretching and angle bending force fields are discussed in the Section~\ref{IPFF},
along with the analytical formulas for the respective potential energies.
Then, the Section~\ref{sstress} contains results concerning the mechanical response
of graphene monolayers on uniaxial tension, hydrostatic tension, and shear stress.
Stress-strain curves and the corresponding elastic parameters are presented.
Finally, the Section~\ref{concl} concludes our work.

\section{In-plane force fields  \label{IPFF} }

We performed first principles', Density Functional Theory (DFT) calculations at the GGA/PBE 
level\cite{functional} for 
an infinite graphene sheet. We used the Quantum-Espresso computer code\cite{quantesp} 
with a ultra-soft pseudopotential~\cite{pseudopotential}
generated by a modified RRKJ approach~\cite{rrkj} which is proven to reproduce 
accurately structural, vibrational and thermodynamic properties of carbon allotropes~\cite{marzari}.
Finally, we  adopted a minimal two-atom unit cell, a 10$\times$10 k-mesh and cutoffs
70 and 560 Ry for the wave functions and charge density respectively.

\begin{figure}
\epsfig{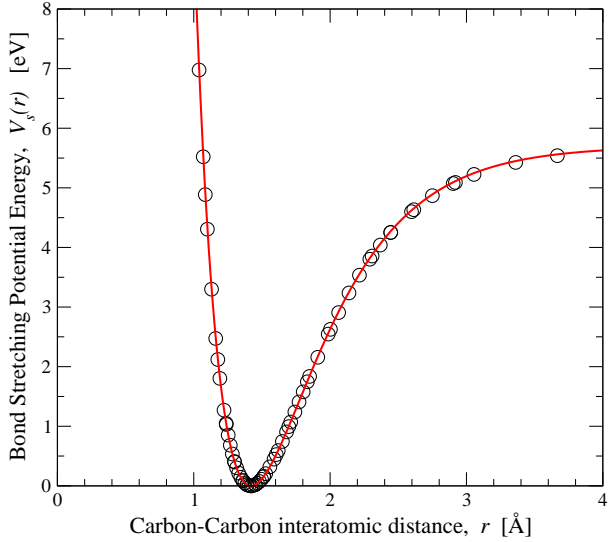}
\caption{  Bond stretching potential energy as a function of the interatomic distance between
bonded carbon atoms in graphene layers. Circles show the results obtained through density functional theory.
Solid line represents the best fit with a Morse potential, Eq. (\ref{bsp}).}
\label{fbspot}
\end{figure}

To find the energy dependence of the bond stretching motion we 
scaled the lattice, thus stretching all C-C bonds without altering any valence angle.
In  Fig.~\ref{fbspot} the calculated energy per bond is shown.
For efficient atomistic simulations, we have fitted the obtained DFT results with
appropriate anharmonic functions, thus providing analytical potentials for the
description of the in-plane molecular mechanics of graphene.
Regarding bond stretching, the Morse potential
\begin{equation}  \label{bsp}
V_s(r) = D \left[ e^{-a(r-r_0)}-1 \right]^2
\end{equation}
has been considered. Through fitting of the numerical DFT results with eq. (1)
, the parameter values
$D=5.7 \; eV$, $a=1.96 \; \AA^{-1}$, $r_0=1.42 \; \AA$ have been obtained (see Fig.~\ref{fbspot} ).

\begin{figure}
\epsfig{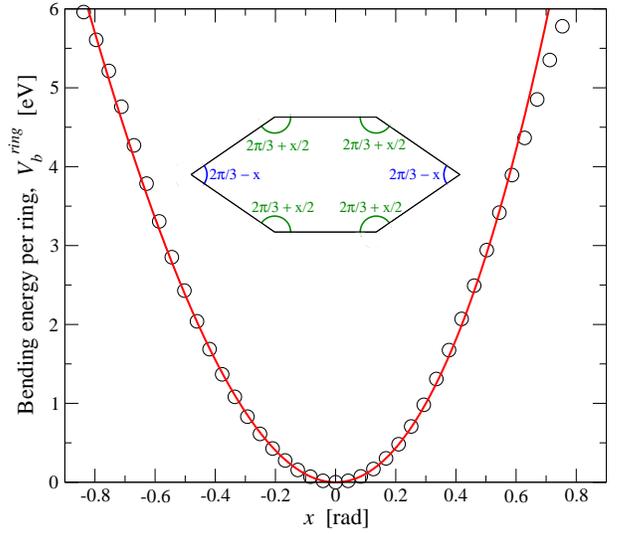}
\caption{  Bending energy per hexagonal ring as a function of the angle deformation shown in the inset.
Circles show the results obtained through density functional theory.
Solid line represents the best fit with Eq. (\ref{bring}), from where the parameters of the angle bending
potential (\ref{abp}) are derived.}
\label{fabpot}
\end{figure}

The angle bending force field is described by a nonlinear potential containing quadratic and cubic terms
\begin{equation}  \label{abp}
V_b(\phi) = \frac{k}{2} \left( \phi-\frac{2\pi}{3} \right)^2 - \frac{k^\prime}{3} \left( \phi-\frac{2\pi}{3} \right)^3.
\end{equation}
To obtain the parameters $k$ and $k^\prime$ of this expression, we consider the deformed graphene
lattice where each  benzene ring is identically deformed as shown in the inset of Fig.~\ref{fabpot};
only the valence angles have been uniformly altered by $x$ or $x/2$, while all bond lengths have their
equilibrium value of 1.42 $\AA$.
In the same figure we show the dependence  of the DFT calculated total energy per ring on the angle-bending
parameter $x$. For such a deformation, considering the angle bending potential energy of Eq.~(\ref{abp}),
the obtained energy per ring is
\begin{equation}  \label{bring}
V_b^{ring} = 2 V_b \left( \frac{2\pi}{3}-x \right) + 4V_b \left( \frac{2\pi}{3}+\frac{x}{2} \right)
=\frac{3}{2}kx^2 + \frac{1}{2} k^\prime x^3,
\end{equation}
Fitting the DFT results of Fig.~\ref{fabpot} with the last expression yields the parameter values
$k=7.0 \; eV/rad^2$ and $k^\prime=4 \; eV/rad^3$ for the angle bending potential, Eq.(\ref{abp}).

We mention that bond stretching and angle bending potentials like those discussed here,
 Eqs. (\ref{bsp}) and (\ref{abp}), have been also presented in Ref. \cite{ffJCP}, but with different
 parameter values:  $D=234.42 \; kcal/mol = 10.1 \; eV$, $a=1.75 \; \AA^{-1}$, $r_0=1.27 \; \AA$,
$k=36.26 \; kcal/mol/rad^2=1.56 \; eV/rad^2$, and $k^\prime=0$ (a linear angle bending potential
has been considered in that work).

\section{Mechanical response of graphene monolayers\label{sstress}}

Stress-strain curves and elastic constants of graphene monolayers have been calculated using
both first principle's method and molecular dynamics (MD) atomistic simulations with the potentials
presented in the previous section.

In the former case, DFT calculations were performed, as previously,
but for a larger rectangular unit cell comprising by 16 atoms and a 6$\times$12 k-mesh.
Strains were applied in one (either the zigzag or armchair, for uniaxial
tensions) or both (for hydrostatic tension) of the two vertical directions of the unit cell. For a given strain level the corresponding unit cell dimensions are fixed while all other structural parameters such as atomic positions and vertical to strain unit-cell
dimension for uniaxial strain, were allowed to relax.
The stresses (forces per length) were calculated as numerical derivatives of the total energy 
with respect to the appropriate length (or area) of the unit cell for given strain.

In order to obtain the mechanical response of graphene at a fixed force, $f$, acting on one
or more of its edges, atomistic simulations have been performed as follows:
we start with the equilibrium structure of graphene, without any external force (i.e. all
first neighboring distances are 1.42 $\AA$ and all valence angles $2\pi/3$). Then, a constant force $f$
is applied at all atoms of the appropriate edge, depending on the case (the force may be
perpendicular or parallel to the edge). Now, the system is out of equilibrium and Newton's
equations of motion are numerically solved, applying a friction term at each atom, to follow
the evolution of the system. Then, the system gradually goes to a new equilibrium, compatible
to the applied forces. In this process, due to the friction the total energy of the system
decreases from its initial value towards a new lower value, corresponding to the new equilibrium
(the deformed state due to the applied stress). The total kinetic energy increases initially from
zero and after some decaying oscillations it gradually vanishes, when the new equilibrium has
been reached. During this evolution, strains are developed in the graphene sheet and, following
a transient oscillatory behavior, they converge to the equilibrium values corresponding to the applied stress.
To avoid finite size effects, the equilibrium strains at the middle of the examined graphene sheets
are recorded. Then the same procedure is repeated for another value of the applied force $f$.
Graphene monolayers consisting of 7482 and 17030 carbon atoms have been simulated, providing
the same results. The friction coefficient was 10 $psec^{-1}$. It has been checked that other values of
the friction coefficient result in the same deformed state, affecting only the transient period and the
number of oscillations needed before reaching the corresponding equilibrium.

\subsection{Uniaxial tension}

Here, atomistic simulations have been performed, as described above, with a constant
force $f$ applied at all atoms belonging to the two opposite edges of graphene.
The force is perpendicular to the corresponding edges and it is directed outwards.
Fixed strains at the two opposite edges of the unit cell have been imposed in the respective
DFT calculations, while the vertical direction has been relaxed to minimize the total energy.
For a given strain, the stress is obtained through the numerical derivative of the total energy 
with respect to the size of the unit cell along the direction where the strains are imposed.
MD (DFT) results are presented for the cases where the tensile force (strain) is applied either
on the zigzag or on the armchair edges of graphene.

\begin{figure}
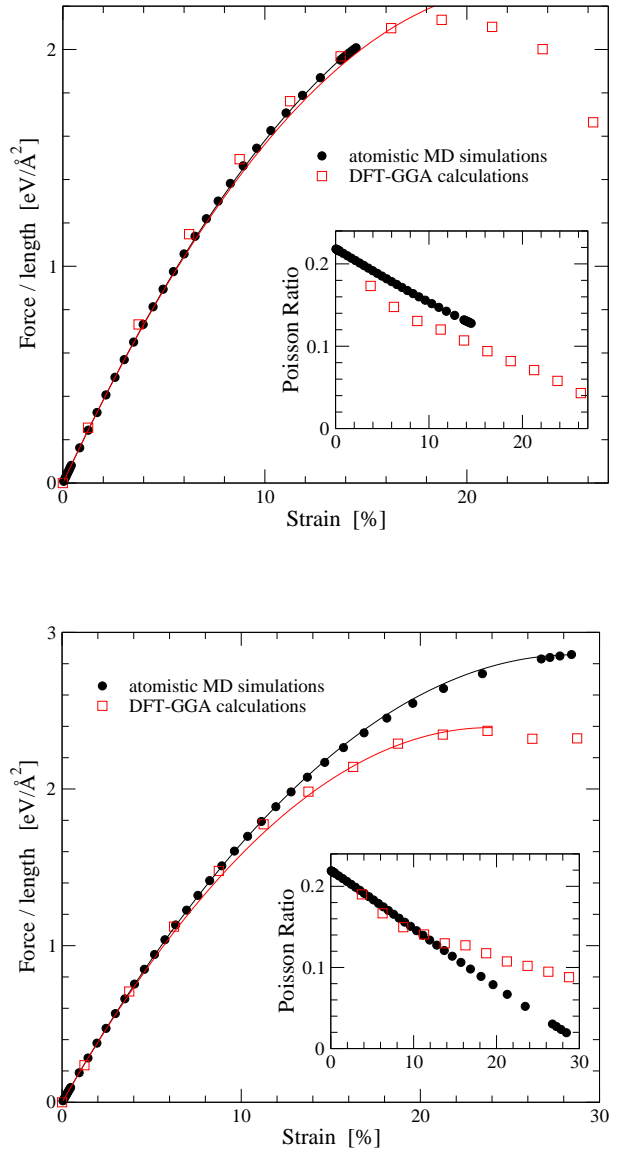

\epsfig{file=Fig3a.eps,width=8.0cm}\\ [1.2cm]
\epsfig{file=Fig3b.eps,width=8.0cm}
\caption{ Stress-strain curves for uniaxial deformations of graphene monolayers.
{\it Top:} The uniaxial tension is applied to graphene's zigzag edges.
{\it Bottom:} The uniaxial tension is applied to graphene's armchair edges.
Filled circles show results obtained through molecular dynamics simulations,
while open squares correspond to density functional theory calculations.
Solid lines represent fits with the nonlinear relation, Eq.~(\ref{nlss}). 
The insets depict the variation of the Poisson ratio with the uniaxial strain.}
\label{funi}
\end{figure}

Fig.~\ref{funi} depicts the force per unit length (the corresponding stress at the edge of  a
two-dimensional material) as a function of the strain (\%) for uniaxial tensions applied on the
graphene's zigzag (top) or armchair (bottom) edges. In the former case, the DFT and MD results
coincide up to a strain of 14-15\%, above of which failure is obtained in the atomistic simulations.
In the latter case, the DFT and MD calculations are almost identical up to about 12-14\% strain,
while failure is obtained at a strain larger than 28\% in the atomistic simulations and a bit
earlier in the first principles method.

The slope of the obtained stress-strain curves in the limit of small strains and stresses
leads to a $2D$ Young modulus $E_{2D}=320 \; N/m$, for uniaxial tensions applied on both
types of graphene edges. Considering the thickness of a graphene monolayer to be $l=0.335 \; nm$
(the interlayer spacing of graphite), the derived effective $3D$ Young modulus is
$E_{eff}=E_{2D}/l=0.96 \; TPa$. The $2D$ intrinsic strengths (the maximum tensile stresses supported
before failure) are $\sigma^z_{2D}=32-34 \; N/m$ for uniaxial forces applied on the zigzag edges
and $\sigma^a_{2D}=39-45 \; N/m$ for uniaxial forces applied on the armchair edges.
The effective $3D$ intrinsic strengths, $\sigma_{eff}=\sigma_{2D}/l$, are around 100 $GPa$
and 120-130 $GPa$, respectively. The failure observed in atomistic MD simulations for stresses
larger than the intrinsic strength, occurs as shown in Fig.~7 of Ref.~\cite{gao}, i.e. when the tensile
force is applied on the armchair (zigzag) edges of graphene, the terminal single layer of atoms on which the forces are applied (the terminal layer of hexagonal atomic rings) breaks and it is disrupted from the system.

The calculated Young modulus and intrinsic strengths are in agreement with available
experimental estimates and in the range of theoretically computed values from other works.
Early measurements of the elastic constants of highly ordered pyrolytic graphite indicated
a value of 1 $TPa$ for the in-plane Young modulus. More recently, indirect experimental estimates,
obtained through nanoindentation measurements of free-standing monolayer graphene~\cite{lee},
result in $2D$ Young modulus  $E_{2D}=340\pm50 \; N/m$ and strength $\sigma_{2D}=42\pm4 \; N/m$,
leading to effective $3D$ values $E_{eff}=1.0\pm0.1 \; TPa$ and $\sigma_{eff}=130 \pm 20 \; GPa$,
respectively. Theoretical predictions for the Young modulus of graphene have been obtained
through a variety of methods. Molecular mechanics simulations yield $E_{eff}=0.95 \; TPa$~\cite{ffJCP},
MD simulations $E_{eff}=1.01 \; TPa$~\cite{zhaoNL},
Monte Carlo calculations $E_{eff}=1.04 \; TPa$ ($E_{2D}=350 \; N/m$)~\cite{fasolino},
tight-binding approximations $E_{eff}=0.91 \; TPa$~\cite{zhaoNL},
and density functional theory methods result in values of $E_{eff}=1.05-1.09 \; TPa$ \cite{liu,yazdi,philipp}.
Intrinsic strengths $\sigma^z_{eff}=110 \; GPa$ and $\sigma^a_{eff}=121 \; GPa$ for forces applied
along the zigzag and armchair edges, respectively, have been calculated in Ref.~\cite{liu}.

The elastic response of monolayer graphene under uniaxial extension can be captured by the
following nonlinear relation
\begin{equation}  \label{nlss}
\sigma_{2D} = E_{2D} \cdot \epsilon + D_{2D} \cdot \epsilon^2,
\end{equation}
where $\epsilon$ is the uniaxial strain and $D_{2D}$ is the third-order elastic modulus which is typically
negative~\cite{lee}. The Young modulus is about 320 $N/m$ for both directions
of uniaxial tension and the $D_{2D}$ is derived through fitting of the numerical data with Eq.~({\ref{nlss}).
When tension is applied on the zigzag edges, the extracted $D_{2D}$ is $-700 \; N/m$
in the DFT  and $-670 \; N/m$ in the MD results. For the armchair direction the obtained $D_{2D}$
values are $-670 \; N/m$ and $-560 \; N/m$, respectively. These values are
in agreement with the experimentally determined value of $-690 \; N/m$~\cite{lee}.

The insets in Fig.~\ref{funi} show the corresponding Poisson's ratio, $\nu$, as a function of the
uniaxial strain. In both directions, a value of $\nu \approx 0.22$ is obtained for small strains,
while the Poisson ratio decreases for larger strains. When the uniaxial tension is applied
to the zigzag edges, the slope of the almost linear decrease of the Poisson ratio is $-0.006/\%strain$
in both MD and DFT calculations. In the other direction, atomistic simulations give a linear decrease
of $\nu$ with a slope of $-0.007/\%strain$, while the DFT calculations coincide with the MD ones for
strains up to about 12\%, but the slope changes for larger strains resulting in an approximate average
slope of $-0.004/\%strain$. Similar behavior has been obtained from the DFT results of
Ref.~\cite{liu}, where $\nu$ starts from a value around 0.19 at small strains and then it almost
linearly decreases up to strains 25-30\%, exhibiting a more abrupt slope in the case of tension
applied on the zigzag edges. Other theoretical estimates give values of $\nu$ around 0.15 at
0~$^\circ K$ \cite{fasolino}, $\nu=0.28$ at 1~$atm$ and 300~$^\circ K$ \cite{ffJCP}, and
$\nu=0.21$ at 300~$^\circ K$ \cite{zhaoNL}.

\subsection{Hydrostatic tension}

In this case, constant forces are applied at all boundary atoms of the graphene sheets studied with MD,
in such a way that the corresponding stress (force per unit length) is the same at both kinds of edges.
Taking into account that the average interatomic distances {\it along} the zigzag and armchair
edges are $\sqrt{3} r_0/2$ and $0.75 r_0$, respectively, the applied forces have different magnitudes
at each type of edge. They are perpendicular to the edges and are directed outwards.
In the DFT calculations, a uniform strain in both vertical directions was applied.
The hydrostatic stress was calculated by numerical differentiation of the energy with respect
to the unit cell's area, at given uniform strains. Applying uniform strain in both directions,
i.e. assuming that the material is isotropic, is a reasonable approximation as it is justified below.

Fig~\ref{fhydro} presents the force per unit length as a function of the relative surface change
$DS/S_0=(S-S_0)/S_0$, where $S_0$ is the initial undeformed surface of the system and $S$
is the final equilibrium surface corresponding to the applied forces.
Graphene supports hydrostatic tension up to more than 30\% relative change of its surface.
The maximum tensile hydrostatic stress before failure is obtained around 31-32 $N/m$, which,
divided by $l$, corresponds to an effective $3D$ stress 90-100 $GPa$.
This value of maximum stress is close to the intrinsic strength of uniaxial tension applied on the
zigzag edge, which is the minimum of the two values of intrinsic strengths corresponding to
the two different directions of uniaxial tension. However, this does not imply that hydrostatic
failure occurs at the zigzag edges by the same mechanism as in the corresponding uniaxial
tension~\cite{marianetti}.
On the contrary, MD simulations indicate a more complicated failure pattern in this case.

\begin{figure}
\epsfig{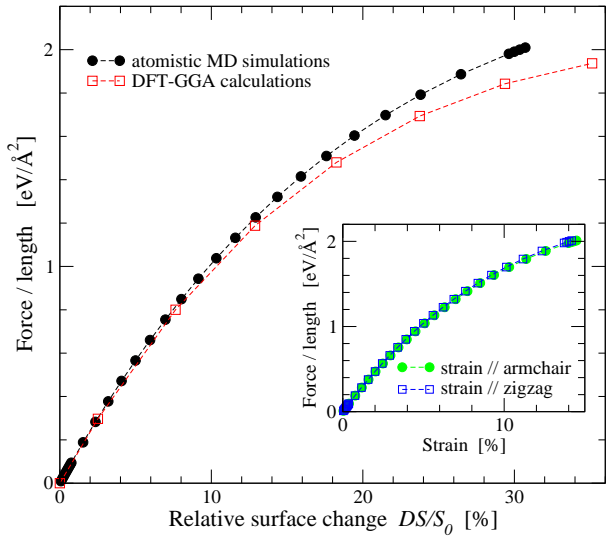}
\caption{  Stress-strain curve for hydrostatic tension deformation of graphene monolayers.
Filled circles (open squares) show results obtained through molecular dynamics simulations
(density functional theory calculations).
The inset depicts the force per unit length as a function of the relative length deformation
along the armchair (filled circles) and the zigzag (open squares) direction, obtained through
the molecular dynamics calculations.
Dashed lines are guides to the eye.}
\label{fhydro}
\end{figure}

In the small strain/stress limit, the slope of the stress-strain curve shown in Fig~\ref{fhydro}
results in a $2D$ bulk modulus $B_{2D}=200 \; N/m$.
This value of the $2D$ bulk modulus is in agreement with estimates of
$B_{2D}\approx 13 \; eV/\AA^2 \approx 200 \; N/m$, obtained from Monte Carlo simulations~\cite{fasolino}.

In order to investigate potential asymmetries in the elastic response of graphene under
hydrostatic load, we show in the inset of Fig~\ref{fhydro} the relation between the applied
force per unit length and the relative extensions along the zigzag and armchair directions,
separately, as obtained from the atomistic simulations. A completely symmetric response
arises, as the strains along the two directions are identical. This confirms the equivalence
between the DFT calculations, obtained by applying uniform strains along the two directions,
and the hydrostatic MD results, where uniform stresses are applied in the two directions.

\subsection{Shear stress}

Shear deformations are investigated merely by MD simulations.To this end, one edge of
graphene has been kept fixed, at its free-of-load equilibrium structure, during
the whole MD process, while constant forces are applied at all atoms of
the opposite edge. The direction of the forces is parallel to the edge on which they apply.
The shear response of both kinds of edges has been calculated through this procedure.

\begin{figure}
\epsfig{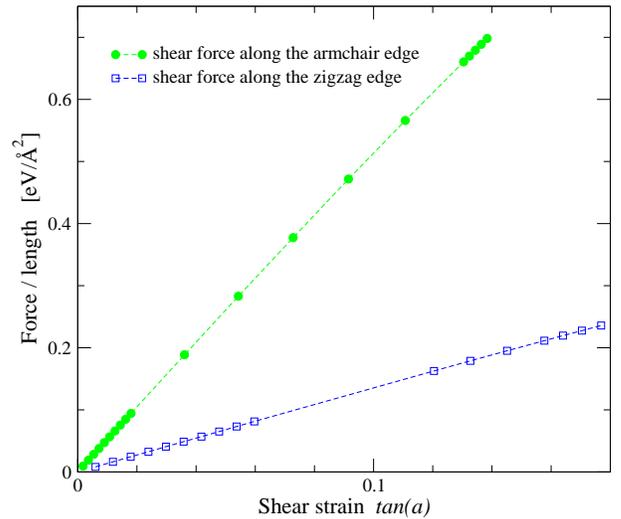}
\caption{ Stress-strain curves for shear deformations of graphene monolayers,
obtained through molecular dynamics simulations.
Filled circles (open squares) show results corresponding to shear forces acting
on the armchair edge (zigzag edge) of graphene.
Dashed lines are guides to the eye. }
\label{fshear}
\end{figure}

Fig. \ref{fshear} shows the shear stress as a function of shear strain (the tangent of the strain
angle at the middle of the examined graphene sheet). A noticeable asymmetry on the shear
response of the zigzag and armchair edges is obtained. The armchair edges can withstand
shear stresses up to around 11 $N/m$, where a maximum shear strain 0.14 is developed.
A significantly smaller maximum shear stress of less than 4 $N/m$ is supported by the zigzag
edges, but with a larger shear strain.

Almost linear stress-strain curves describe the shear deformations of both zigzag and armchair
edges of graphene. The obtained $2D$ shear moduli are  $G^a_{2D} = 84 \; N/m$ for shear
stresses applied on the armchair edge and $G^z_{2D} =22 \; N/m$ for shear applied on the
zigzag edge. The effective $3D$ shear moduli $G_{eff}=G_{2D}/l$ are 250 $GPa$ and 65 $GPa$,
respectively. The asymmetric shear response of the two kinds of edges result in a difference
by a factor of about 4 of the corresponding elastic moduli. A higher value of shear modulus in
graphene, around 150 $N/m$, has been obtained from Monte Carlo simulations~\cite{fasolino}.

\subsection{Effect of the anharmonicity of force fields}

As it has been mentioned in the Section~\ref{intro}, one of the advantages of atomistic simulations,
when analytical force fields are available, is that the influence of several factors can be explored
by altering or eliminating respective potential energy terms. Here, we examine the effect of
nonlinear interactions on the uniaxial tensile response of graphene, since both bond stretching
and angle bending potentials, Eqs. (\ref{bsp}) and (\ref{abp}), presented in this work are anharmonic.

Using respective MD simulations, results have been obtained for the cases that
(i) the angle bending potential is linearized by ignoring the cubic term (setting $k^\prime=0$)
in Eq.~(\ref{abp}) and (ii) both angle bending and bond stretching potentials are linearized; i.e.
additionally to the change mentioned previously, the Morse potential is also linearized through
the substitution of Eq.~(\ref{bsp}) by its harmonic approximation $V_s^{linear}=Da^2(r-r_0)^2$.

\begin{figure}
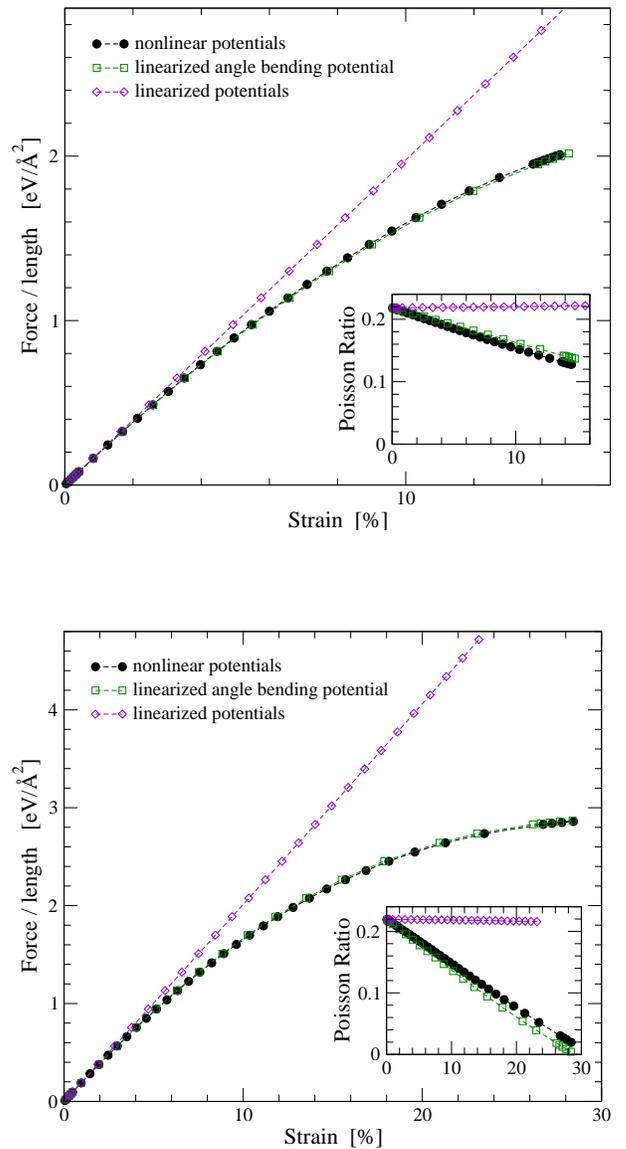

\epsfig{file=Fig6a.eps,width=8.0cm}\\ [1.25cm]
\epsfig{file=Fig6b.eps,width=8.0cm}\\
\caption{ The effect of the anharmonicity of the potentials on the uniaxial deformations of
graphene monolayers.  Filled circles show results using the fully nonlinear potentials of
Eqs. (\ref{bsp}) and (\ref{abp}), open squares correspond to linearized angle bending potential
and nonlinear bond stretching potential, while open diamonds correspond to completely
linearized potentials. All the results are obtained using molecular dynamics simulations.
{\it Top:} The uniaxial deformation is applied to graphene's zigzag edges.
{\it Bottom:} The uniaxial deformation is applied to graphene's armchair edges.
The insets depict the corresponding results for the Poisson ratio as a function
of the uniaxial strain. Dashed lines are guides to the eye. }
\label{flinear}
\end{figure}

Fig. \ref{flinear} compares the stress-strain curves on uniaxial tension obtained in the cases (i) and (ii)
with the corresponding results shown in Fig.~\ref{funi}, where the anharmonic potentials of
Eqs. (\ref{bsp}) and (\ref{abp}) have been used. The insets present these comparisons regarding
the strain dependence of the Poisson ratio. It can be seen that the nonlinearity of the angle bending
potential has not any noticeable effect on the calculated mechanical response.
Instead, the response is drastically different when harmonic bond stretching potentials are considered
too. As expected, in the latter case a linear stress-strain relation is obtained (deviating from the
nonlinear response for strains above a few per cent) and no failure is present. Further, the Poisson
ratio shows no dependence on the strain, in marked contrast to the behavior observed for nonlinear
bond stretching force fields.

\section{Conclusions\label{concl}}

Empirical bond stretching and angle bending potentials have been presented, suitable for
atomistic simulations of graphene. The potentials have been derived using methods from first
principles and are expected to accurately represent the respective interactions in graphene.
Appropriate nonlinear functions that analytically describe the numerically obtained potentials
are provided, for efficient use in molecular mechanics calculations.

The mechanical response of graphene under various loads is examined, through both molecular
dynamics simulations, using the calculated empirical force fields, and density functional theory
calculations. Stress-strain curves have been presented for uniaxial tensions or shear stresses
applied on both armchair and zigzag edges of graphene, as well as for hydrostatic tension.
The dependence of Poisson ratio on strain is also calculated in the case of uniaxial tensions.
Corresponding elastic moduli, such as Young modulus, two-dimensional bulk modulus, or
shear moduli, and maximum values of supported stresses or strain before failure
are discussed and compared with experimental estimates, when available, and other
theoretical results.

Finally, the effect of the nonlinearity of the potentials on the uniaxial tensile response of graphene
is considered. We find that the anharmonicity in the angle bending potential plays no role
in the examined mechanical response, but, instead, nonlinear terms of the bond stretching
potential crucially affect the obtained elastic behavior and, thus, should be necessarily taken into
account.

{\it Acknowledgements}   We acknowledge support from the Thales Project
GRAPHENECOMP, co-financed by the European Union (ESF) and the Greek
Ministry of Education (through E$\Sigma\Pi$A program).


\vspace{-0.5cm}

\begin{thebibliography}{39}


\bibitem{Zhao}   Q. Zhao,  M.B. Nardelli, J. Bernholc, Phys. Rev. B {\bf 65}, 144105 (2002).

\bibitem{Kudin}  K.N. Kudin, G.E. Scuseria, B.I. Yakobson, Phys. Rev. B {\bf 64}, 235406 (2001).

\bibitem{Odegard}  G.M. Odegard, T.S. Gates, L.M. Nicholson, K.E. Wise,
Compos. Sci. Technol. {\bf 62}, 1869 (2002).

\bibitem{Huang} Y. Huang, J. Wu, K.C. Hwang, Phys. Rev. B {\bf 74}, 245413 (2006).

\bibitem{Meo}  M. Meo, M. Rossi, Compos. Sci. Technol. {\bf 66}, 1597 (2006).

\bibitem{liu} F.  Liu, P.  Ming, J. Li,  Phys. Rev. B {\bf 76}, 064120 (2007).

\bibitem{Cadelano}  E. Cadelano, P.L. Palla, S. Giordano, L. Colombo,
Phys. Rev. Lett. {\bf 102}, 235502 (2009).

\bibitem{fasolino}  K.V. Zakharchenko, M.I. Katsnelson, A. Fasolino,
Phys. Rev. Lett. {\bf 102}, 046808 (2009).

\bibitem{zhaoNL}  H. Zhao, K. Min, N.R. Aluru,  Nano Lett. {\bf 9}, 3012 (2009).

\bibitem{Pour}  A.Sakhaee-Pour, Solid State Communications {\bf 149}, 91 (2009).

\bibitem{Ni}  Z. Ni, H. Bu, M. Zou, H. Yi, K. Bi, Y. Chen, Physica B {\bf 405}, 1301 (2010).

\bibitem{Tsai}  J.-L. Tsai, J.-F. Tu, Materials and Design {\bf 31}, 194 (2010).

\bibitem{lee}  C. Lee, X. Wei, J.W. Kysar, J. Hone, Science  {\bf 321}, 385 (2008).

\bibitem{Tsouk}  G. Tsoukleri, J. Parthenios, K. Papagelis, R. Jalil, A.C. Ferrari, A.K. Geim,
K.S. Novoselov, C. Galiotis, Small {\bf 21}, 2397 (2009).

\bibitem{OF1}  O. Frank, G. Tsoukleri, I. Riaz, K. Papagelis, J. Parthenios, A.C. Ferrari, A.K. Geim,
K.S. Novoselov, C. Galiotis, Nature Communications 2:255 doi: 10.1038/ncomms1247 (2011).

\bibitem{zabel}   J. Zabel, R.R. Nair, A. Ott, T. Georgiou, A.K. Geim, K.S. Novoselov,
C. Casiraghi, Nano Lett. {\bf 12}, 617 (2011).

\bibitem{OF2}  O. Frank, M. Bouoa, I. Riaz, R. Jalil, K.S. Novoselov, G. Tsoukleri, J. Parthenios,
L. Kavan, K. Papagelis, C. Galiotis, Nano Lett. {\bf 12}, 687 (2012).

\bibitem{KNF} G. Kalosakas, K.L. Ngai, S. Flach, Phys. Rev. E {\bf 71}, 061901 (2005). 

\bibitem{functional} J.P. Perdew, K. Burke, M. Ernzerhof, Phys. Rev. Lett. {\bf 77},
3865 (1996). 

\bibitem{quantesp}  P. Giannozzi, S. Baroni {\it et al.}, J. Phys.: Cond. Matt. {\bf 21}, 395502 (2009).

\bibitem{pseudopotential} A. Dal Corso, C.pbe-rrkjus.upf, 
http://www.quantum-espresso.org/wp-content/uploads/upf\_files/C.pbe-rrkjus.UPF

\bibitem{rrkj}  A.M. Rappe, K.M. Rabe, E. Kaxiras, J.D. Joannopoulos,
Phys. Rev. B {\bf 41}, R1227 (1990).

\bibitem{marzari}  N. Mounet, N. Marzari, Phys. Rev. B {\bf 71}, 205214 (2005).

\bibitem{ffJCP}   D. Wei, Y. Song, F. Wang, J. Chem. Phys. {\bf 134}, 184704 (2011).

\bibitem{gao}  Y. Gao, P. Hao,  Physica E {\bf 41}, 1561 (2009).

\bibitem{jap70}  O.L. Blakslee, D.G. Proctor, E.J. Seldin, G.B. Spence, T. Weng,
J. Appl. Phys. {\bf 41}, 3373 (1970).

\bibitem{yazdi}  C.D. Zeinalipour-Yazdi, C. Christofides, J. Appl. Phys. {\bf 106}, 054318 (2009).

\bibitem{philipp}  P. Wagner, C.P. Ewels, V.V. Ivanovskaya, P.R. Briddon, A. Pateau, B. Humbert,
Phys. Rev. B {\bf 84}, 134110 (2011).

\bibitem{marianetti}  C.A. Marianetti, H.G. Yevick, Phys. Rev. Lett. {\bf 105}, 245502 (2010).

\end{thebibliography}
\end{document}